\documentclass[aps,pra,twocolumn,byrevtex,nofootinbib,
				showpacs,floatfix,
				superscriptaddress]{revtex4-1}

\usepackage{graphicx}
\usepackage{amsmath}
\usepackage{mathrsfs}
\usepackage{bm}
\usepackage{amssymb}
\usepackage{microtype}
\usepackage{hyperref}
\hypersetup{colorlinks=true,linkcolor=blue,citecolor=blue,urlcolor=blue}
\newcommand{\la}{\langle}
\newcommand{\ra}{\rangle}

\begin{document}

\title{Experimental quantum tomography assisted by multiply symmetric states in higher dimensions}

\author{D. Mart\'inez}
\affiliation{Departamento de F\'{\i}sica, Universidad de Concepci\'on, 160-C Concepci\'on, Chile}
\affiliation{Millennium Institute for Research in Optics, Universidad de Concepci\'on, 160-C Concepci\'on, Chile}

\author{M. A. Sol\'is-Prosser}
\email{msolisp@udec.cl}
\affiliation{Departamento de F\'{\i}sica, Universidad de Concepci\'on, 160-C Concepci\'on, Chile}
\affiliation{Millennium Institute for Research in Optics, Universidad de Concepci\'on, 160-C Concepci\'on, Chile}

\author{G. Ca\~nas}
\affiliation{Departamento de F\'isica, Universidad del B\'io-B\'io, Collao 1202, Casilla 5C, Concepci\'on, Chile}

\author{O. Jim\'enez}
\affiliation{Centro de \'Optica e Informaci\'on Cu\'antica, Facultad de Ciencias,
Universidad Mayor, Chile}

\author{A. Delgado}
\affiliation{Departamento de F\'{\i}sica, Universidad de Concepci\'on, 160-C Concepci\'on, Chile}
\affiliation{Millennium Institute for Research in Optics, Universidad de Concepci\'on, 160-C Concepci\'on, Chile}

\author{G. Lima}
\affiliation{Departamento de F\'{\i}sica, Universidad de Concepci\'on, 160-C Concepci\'on, Chile}
\affiliation{Millennium Institute for Research in Optics, Universidad de Concepci\'on, 160-C Concepci\'on, Chile}

\begin{abstract}
High-dimensional quantum information processing has become a mature field of research with several different approaches being adopted for the encoding of $D$-dimensional quantum systems. Such progress has fueled the search of reliable quantum tomographic methods aiming for the characterization of these systems, being most of these methods specifically designed for a given scenario. Here, we report on a new tomographic method based on multiply symmetric states and on experimental investigations to study its performance in higher dimensions. Unlike other methods, it is guaranteed to exist in any dimension and provides a significant reduction in the number of measurement outcomes when compared to standard quantum tomography. Furthermore, in the case of odd dimensions, the method requires the least possible number of measurement outcomes. In our experiment we adopt the technique where high-dimensional quantum states are encoded using the linear transverse momentum of single photons and are controlled by spatial light modulators. Our results show that fidelities of $0.984\pm0.009$ with ensemble sizes of only $1.5\times10^5$ photons in dimension $D=15$ can be obtained in typical laboratory conditions, thus showing its practicability in higher dimensions.
\end{abstract}

\date{\today}

\maketitle

\section{Introduction}

The generation, manipulation and measurement of high-dimensional quantum systems (qudits) are important theoretical and experimental research subjects in quantum information science.  This is motivated, in part, because certain fundamental features of quantum mechanics such as, for instance, quantum contextuality~\cite{Specker1960,Kochen1975,Canas2014}, cannot be tested with 2-dimensional quantum systems. The use of high-dimensional quantum systems also leads to improvements in several entangled based quantum information protocols since, in this case, some Bell inequalities exhibit increased robustness against noise~\cite{Kaszlikowski2000,Collins2002}, and tolerate lower detection efficiencies for closing the detection loophole~\cite{Vertesi2010}. Last, due to the larger amount of information that can be encoded in single qudits, the performance of several protocols in quantum communications~\cite{Brukner2002,Bruss2002,Cerf2002,Durt2003,Nikolopoulos2006,Ali-Khan2007}  and quantum computation~\cite{OLeary2006,Lanyon2009,Luo2014,Gedik2015,Tonchev2016} is enhanced when they are employed. Typically, photonic platforms are used as testbed experiments to study quantum information processing in higher dimensions because different degrees of freedom of single photons can be efficiently used to encode the qudits. For instance, one can resort to the orbital angular momentum~\cite{Mair2001,Langford2004,Jack2009,Dada2011,Agnew2014}, frequency~\cite{Law2000,Bernhard2013,Kues2017}, time bin~\cite{Takesue2009a}, path~\cite{Rossi2009}, and the transverse position/momentum~\cite{Neves2004,Neves2005} encoding methods.

Quantum tomography (QT) is a collection of methods that makes possible the estimation of unknown quantum states~\cite{ParisBook2004}. Today, QT has become a standard tool for the quality assessment of the generation of quantum states~\cite{Breitenbach1997a,Ourjoumtsev2007}, the implementation of quantum processes~\cite{Chuang1997,Chuang1998,Mohseni2008}, and the performance of quantum devices~\cite{DAriano2002,Titchener2018}. Quantum tomographic methods provide an estimate of the unknown state from the outcomes of measurements carried out on an ensemble of identically, independently prepared systems. Finite statistics effects and unavoidable experimental errors require the postprocessing of the experimentally acquired data by means of statistical inference methods such as, for instance, maximum likelihood estimation~\cite{Hradil1997,James2001,James2001a,Rehacek2001} or bayesian inference~\cite{Jones1991,Slater1995,Buzek1998,Schack2001,Huszar2012,Kravtsov2013,Struchalin2016,Granade2016,Goncalves2018}. Traditionally, the total number of measurement outcomes is considered as a resource. Thus, there is a search for QT methods relying on a smaller number of measurement outcomes~\cite{Cramer2010,Gross2010,Goyeneche2015,PearsStefano2017}. Standard quantum tomography for a single qudit is based on the measurement of a $D$-dimensional representation of the $D^2-1$ generators of the SU($D$) group, which leads to a total number of measurement outcomes of $2D^2-D$~\cite{Thew2002}. This number can be reduced to $D^2+D$ with quantum tomography based on mutually unbiased bases (MUB)~\cite{Wootters1989}. The existence of MUB has been proven when the dimension $D$ is an integer power of a prime number~\cite{Bandyopadhyay2002,Durt2010}. Otherwise, the existence of mutually unbiased bases is still an open problem. A further reduction can be achieved with quantum tomography based on  a symmetric informationally complete (SIC) positive-operator valued measure (POVM), which consists of $D^2$ sub-normalized projectors~\cite{Renes2004}. This is the smallest number of measurement outcomes to estimate unknown quantum states. Numerical studies have indicated the existence of this class of measurements in all dimensions $D\le67$ and algebraic demonstrations are available in dimensions $D=4,\dots,15,19,24, 35$, and $48$~\cite{Scott2010}. Unfortunately, a dimension-independent demonstration is still missing.

Here, we propose and experimentally test a new quantum tomographic method, which is based on the measurement of an informationally complete POVM formed by sub-normalized projectors onto multiply symmetric states~\cite{Paiva-Sanchez2010}. These are constructed by applying products of integer powers of unitary transformations on a fixed fiducial quantum pure state. Unlike SIC-POVM and MUB, our tomographic method can be constructed in any finite dimension. Furthermore, in the case of odd dimensions, the POVM has $D^2$ sub-normalized projectors and thus it requires the smallest number of measurement outcomes to estimate unknown quantum states. In the case of even dimensions, the POVM has $3D^2/2$ measurement operators, which is a significative reduction from the case of standard tomography. A first estimate of the unknown state can be obtained by linear inversion, which does not introduce bias~\cite{Schwemmer2015}. The numerical stability of this process can be improved at a great extent by a suitable choice of the fiducial state. This also contributes to speed up the rate of convergence in the postprocessing of the experimentally acquired data.  Our experimental implementation is based on the encoding of $D$-dimensional quantum states onto the linear transverse momentum of single photons. These are created by defining $D$ different propagation paths available for the photon transmission at diffractive apertures addressed on spatial light modulators (SLM)~\cite{Lima2009f}. A second set of SLMs allows one to project the $D$-dimensional state onto any other fixed $D$-dimensional state \cite{Lima2011}. The use of SLMs for preparation~\cite{Lima2009f,Lima2011,Solis-Prosser2013,Varga2014,Varga2017} and measurement of these so-called spatial qudits has been extensively explored for quantum information tasks such as QKD~\cite{Etcheverry2013}, Bell-type nonlocality and noncontextuality tests~\cite{Dada2011,Canas2014a,Arias2015,Canas2016}, and quantum tomography~\cite{Pimenta2010,Lima2011,Pimenta2013,Goyeneche2015}, among others~\cite{Solis-Prosser2017,Aguilar2018,Martinez2018}. We test our tomographic method in dimensions 6 and 15 reaching fidelities of $0.998$ and $0.984$ with ensemble sizes of only $6\times10^4$ and $1.5\times10^5$, respectively. Experiments performed with similar optical setups have achieved lower fidelities of $0.96$ for dimensions $6$ and $7$, $0.985$ for dimension $8$, and $0.887$ for dimension $10$~\cite{Lima2011,Bent2015,Goyeneche2015} while resorting to larger ensembles of detected photons. Thus, our results demonstrate the practicability of our method in higher dimensions.

This article is organized as follows: In Sec. \ref{sec:theory},  we introduce the theoretical background and formulate our tomographic method. In Sec.~\ref{sec:setup}, we introduce the experimental setup and analyze the results provided by the experimental realization of our method. In Sec.~\ref{sec:conclusion}, we summarize, comment on possible extensions to the multipartite case, and conclude.

\section{Theory \label{sec:theory}}

In this section we briefly recall the general notion of multiply symmetric states. Thereafter, we study a particular family of multiply symmetric states and build the tomographic method upon it. We solve explicitly the inversion problem and provide a simple analytical expression relating the experimentally acquired data, the measurement settings, and the estimate of the unknown state.

\subsection{Multiply symmetric states}

In general, states $|\psi_{k_1,k_2,\dots,k_M}\rangle$ are said to be multiply symmetric if they can be written as~\cite{Barnett2001}
\begin{align}
	|\psi_{k_1,k_2,\dots,k_M}\rangle = U_1^{k_1} U_2^{k_2}\cdots U_M^{k_M}|\psi_{0,0,\dots,0}\rangle,
\end{align}
where $k_j = 0,\dots,N_j-1$, $|\psi_{0,0,\dots,0}\rangle$ is the fiducial state of the set, and $U_j$ are unitary transformations that satisfy $U_j^{N_j}=\mathbb{I}$ (for every $j$), where $\mathbb{I}$ is the identity operator acting onto the Hilbert space of a single qudit. We will limit ourselves to the case of $M=3$. Thus, we define the constant matrices
\begin{align}
	\mathcal{X} =& \sum_{k=0}^{D-1} |k\oplus 1\rangle\langle k|, \label{eq:X}\\
	\mathcal{Z} =& \sum_{k=0}^{D-1} e^{2\pi i k/D}|k\rangle\langle k|, \label{eq:Z} \\
	\mathcal{V} =& \sum_{k=0}^{\bm{\kappa}-1} |k\rangle\langle k| - i\sum_{k=\bm{ \kappa}}^{D-1} |k\rangle\langle k|, \label{eq:V}
\end{align}
where $D$ is the dimension of the Hilbert space and $\bm{\kappa}=[\![ D/2 ]\!]$. These matrices represent, respectively, the shift operator ($\mathcal{X}$), the clock operator ($\mathcal{Z}$) and an additional phase-only transform ($\mathcal{V}$) with diagonal entries $v_k=\langle k|\mathcal{V}|k\rangle$ that adopt values of $1$ and $-i$. The symbol $\oplus$ in Eq.\thinspace(\ref{eq:X}) denotes addition mod($D$). By using the operators $\mathcal{X}$, $\mathcal{Z}$, and $\mathcal{V}$, we now define a set of multiply symmetric states $\{|\alpha_{\ell,m,j}^{\,}\rangle\}$ given by
\begin{align}
	|\alpha_{\ell,m,j}^{\,}\rangle = \mathcal{V}^\ell \mathcal{X}^{m}\mathcal{Z}^{j}|\alpha_0\rangle = \sum_{k=0}^{D-1} a_k v_{k\oplus m}^\ell e^{2\pi i jk/D}|k\oplus m\rangle  , \label{eq:multisym}
\end{align}
where  ${\ell=0,1,2,3}$, ${m=0,\dots,D-1}$, and ${j=0,\dots,D-1}$. The fiducial state is a pure quantum state $| \alpha_0\rangle=\sum_{k=0}^{D-1}a_k|k\rangle$, whose coefficients fulfill the normalization condition $\sum_{k=0}^{D-1}|a_k|^2=1$.

\subsection{Essential subsets of states}
\par Let us now consider a physical system described by an unknown $D$-dimensional quantum state $\rho$. In 2010, Paiva-Sanchez and coworkers~\cite{Paiva-Sanchez2010} studied quantum state tomography assisted by a basis $\mathcal{B}_0(\alpha)$ of $D$ equidistant states, which they denoted as $|\alpha_j\rangle$. These states are such that the inner product between them is given by $\langle\alpha_{j}|\alpha_{j'}\rangle=\alpha$ ($j> j'$), where $\alpha$ is a fixed constant.
Additional $D-1$ bases $\mathcal{B}_s(\alpha)$ are constructed by applying $\mathcal{X}_{\,}^s$ on the elements of $\mathcal{B}_0(\alpha)$. This amounts for a total of $D^2$ measurements. Additionally, they report a strange behavior that depends on the dimension of the Hilbert space where the state belongs to. In summary, odd dimensions require the aforementioned $D^2$ measurements only, whereas even dimensions require additional measurements attainable by applying $\mathcal{V}$ on the elements of each $\mathcal{B}_s(\alpha)$, which leads to a total of $3D^2/2$ measurements.

\par The form of the equidistant states used in Ref.~\cite{Paiva-Sanchez2010} for this purpose resembles the one of Eq.\thinspace(\ref{eq:multisym}). Nevertheless, an  analysis of the computations of Ref.~\cite{Paiva-Sanchez2010} indicates that a similar mathematical procedure allows us to accomplish such tomographic process regardless of the fiducial state used, that is, states $|\alpha_{j}\rangle$ do not need to be equidistant. Thus, we resorted to multiply symmetric states for such goal. If $D$ is an odd number, quantum state tomography can be performed by measuring on projectors of the form $|\alpha_{0,m,j}\rangle\langle \alpha_{0,m,j}|$, with $\ell=0$ and for every $m$ and $j$ ranging from $0$ to $D-1$. For even dimensions, we must also consider $\ell=1$, with $j=0,\dots,D-1$ and $m=0,\dots,D/2-1$ for these additional measurements.

\begin{table}[!t]
\begin{ruledtabular}
\begin{tabular}{ccc}
Value of $s$ & $K_s$ odd dimension & $K_s$ even dimension \\ \hline
	$0\leqslant s\leqslant \bm{\kappa}-1$ & $D$ & $2D$ \\
	$\bm{\kappa}\leqslant s\leqslant D-1$ & $D$ & $D ~$ \\
	$D\leqslant s\leqslant D-1+\bm{\kappa}$ & Does not apply & $2D$
\end{tabular}
\end{ruledtabular}	
\caption{\label{tab:Ks} Values of $K_s$ depending on the values of $s$ and dimension, where $\bm{\kappa} = [\![D/2]\!]$.}
\end{table}

\par In this context, a simpler mathematical description can be obtained by resorting to two subscripts only, regardless of the dimension. Thus, we define
\begin{align}
	|\alpha_{sj}^{\,}\rangle =& \mathcal{V}^{\lfloor s/D \rfloor} \mathcal{X}^{s}\mathcal{Z}^{j}|\alpha_0\rangle, \nonumber \\
		=& \sum_{k=0}^{D-1} a_k v_{k\oplus s}^{(s)}e^{2\pi i jk/D}|k\oplus s\rangle, \label{eq:multisym2}
\end{align}
where ${v_{k}^{(s)} = \langle k| \mathcal{V}^{\lfloor s/D \rfloor} |k\rangle}$, ${j=0,\dots,D-1}$, ${s=0,\dots,s_{\sf max}-1}$, and
\begin{align}
	s_{\sf max} &= \begin{cases}
		D,	~~~~&\text{if $D$ is odd,} \\
		\dfrac{3D}{2},~~~~&\text{if $D$ is even.}
	\end{cases}
\end{align}
Thus, the set of states $\{|\alpha_{sj}\rangle \}$ in odd dimensions is still a complete set of multiply symmetric states under transformations $\mathcal{X}$ and  $\mathcal{Z}$, as seen from Eq.\thinspace(\ref{eq:multisym}). For even dimensions, on the other hand, this set encompasses a \emph{subset} of the multiply symmetric states under the action of $\mathcal{X}$, $\mathcal{Z}$, and $\mathcal{V}$. Despite the different behavior exhibited by states $|\alpha_{sj}\rangle$ as defined here, they allow one to construct POVMs. Indeed, we may define
\begin{align}
	\Pi_{sj} = \frac{1}{K_s}|\alpha_{sj}\rangle\langle\alpha_{sj}|,~~~~~~~ \sum_{s=0}^{s_{\sf max}-1}\sum_{j=0}^{D-1} \Pi_{sj} = \mathbb{I}, \label{eq:POVM}
\end{align}
where the values of $K_s$ are given in Table~\ref{tab:Ks}. This POVM will be useful for tomographic and post-processing purposes.

\subsection{Tomography using multiply symmetric states}
\par Let us define the matrix $\mathcal{P} = \sum_{s,j}p_{sj}|s\rangle\langle j|$, where $p_{sj} = {\rm tr}(\rho \Pi_{sj})$. Explicitly, 
\begin{align}
	\mathcal{P}
		=& \sum_{s=0}^{s_{\sf max}-1}\sum_{j=0}^{D-1}\Bigg( \sum_{l,m=0}^{D-1} \frac{ a^{\ast}_l a_m}{K_s} e^{2\pi i (m-l)j/D} \nonumber \\
			&\hspace{2.6cm}\times v_{l\oplus s}^{(s) \ast} v_{m\oplus s}^{(s)} \rho_{l\oplus s,m\oplus s} \Bigg) |s\rangle\langle j| . \label{Pmatrix2}
\end{align}
This matrix contains the experimental probabilities that can be found by taking the completeness relation of Eq.~(\ref{eq:POVM}) into consideration. So, if $n_{sj}$ is the number of registered counts when $\Pi_{sj}$ is measured, then every probability can be experimentally estimated as ${p_{sj} = n_{sj}/\sum_{t,k} n_{tk}}$. Afterwards, a right-Fourier transformed probability matrix $\mathcal{\widetilde{P}}$ can be defined as $\mathcal{P\cdot F}$,
where $\mathcal{F}= \frac{1}{\sqrt{D}}\sum_{l,m=0}^{D-1} e^{2\pi i lm/D}|l\rangle\langle m|.$ Explicitly,
\begin{align}
	\mathcal{\widetilde{P}}
		=&  \sum_{k=0}^{D-1}\Bigg[\sum_{s=0}^{s_{\sf max}-1} \frac{\sqrt{D}}{K_s} |s\rangle \label{eq:Ptilde1} \\
		&\hspace{0.75cm}\times \Bigg(\sum_{q=0}^{D-1}a^{\ast}_{q\ominus s\oplus k} a_{q\ominus s} v_{q\oplus k}^{(s) \ast} v_{q}^{(s)} \langle q|\Bigg)|\bm{\vec\rho}_k\rangle \Bigg]\langle k| ,\nonumber
\end{align}
where $|\bm{\vec\rho}_m\rangle$ denotes the $m$-th diagonal of $\rho$, given by\footnote{Throughout this document, notation $|\bm{\vec{a}}\rangle$ will refer to a purely mathematical vector $\bm{\vec{a}}$ that does not represent any physical state. However, Dirac notation is used for comfortability. }
\begin{align}
	|\bm{\vec\rho}_m\rangle = \sum_{q=0}^{D-1}\rho_{q\oplus m,q}|q\rangle.
\end{align}
For convenience, we will define ancillary vectors
\begin{align}
	|\bm{\vec{\xi}}_{sk}\rangle
		=& \Big( \mathcal{X}^{s - k}|\alpha_0\rangle \Big) \circ \Big( \mathcal{X}^{s}|\alpha_0\rangle^{\ast} \Big) \circ \Big(\mathcal{X}^{-k}|\bm{\vec{v}}_{s}\rangle \Big) \circ |\bm{\vec{v}}_{s}\rangle^{\ast}, \label{eq:Xisk}
\end{align}
where ``$\circ$'' denotes the Hadamard product between matrices, and
\begin{align}
	|\bm{\vec{v}}_{s}\rangle = \sum_{r=0}^{D-1}v_{r}^{(s)}|r\rangle = \text{diag}\left(\mathcal{V}^{\lfloor s/D\rfloor}\right).
\end{align}
Consequently, the right-transformed probability matrix can be compactly written as
\begin{align}
	\widetilde{\mathcal{P}}
		=& \sum_{k=0}^{D-1} \mathscr{G}_k | \bm{\vec{\rho}}_{k}\rangle \langle k|, \label{eq:Ptilde2}
\end{align}
where matrix $\mathscr{G}_k$ is given by
\begin{align}
	\mathscr{G}_k = \sum_{s=0}^{s_{\sf max}-1} \frac{\sqrt{D}}{K_s}|s\rangle\langle \bm{\vec{\xi}}_{sk} |. \label{eq:Gk}
\end{align}

Now, it is possible to construct the density operator $\rho$ by rearranging its components in a vector $\vec{\rho}=\text{vec}(\rho)$ (see Appendix~\ref{app:vec}), which is computed according to
\begin{align}
	\bm{\vec{\Delta}}_\rho =& \sum_{k=0}^{D-1} |k\rangle\otimes|\bm{\vec{\rho}}_k\rangle, \\
	\vec{\rho} =& \bm{\mathcal{S}}_\textsc{swap}\bm{\mathcal{X}}\cdot \bm{\vec{\Delta}}_\rho, \label{eq:vec_rho_1}
\end{align}
where $\otimes$ represents the Kronecker product between matrices, $\bm{\vec{\Delta}}_\rho$ is a $D^2$-dimensional vector containing the diagonals of $\rho$---given by $|\bm{\vec{\rho}}_k\rangle$---stacked on top of each other, $\bm{\mathcal{S}}_\textsc{swap}$ is a $D^2\times D^2$-matrix that acts as $\bm{\mathcal{S}}_\textsc{swap}( |j\rangle\otimes|k\rangle) = |k\rangle \otimes |j\rangle,$ and
\begin{align}
	\bm{\mathcal{X}} =& \left(\sum_{m=0}^{D-1}\mathcal{X}^{m}\otimes |m\rangle\langle m|\right).
\end{align}

\par Finally, after taking Eqs.~(\ref{eq:Ptilde1}), (\ref{eq:Xisk}), (\ref{eq:Ptilde2}), (\ref{eq:Gk}), and~(\ref{eq:vec_rho_1}) into account, the components of $\rho$ can be isolated by computing
\begin{align}
	\bm{\mathcal{G}} =& \sum_{m=0}^{D-1} |m\rangle\langle m|\otimes \mathscr{G}_m \nonumber \\
		=& \sum_{m=0}^{D-1}\sum_{s=0}^{s_{\sf max}-1} \frac{\sqrt{D}}{K_s} |m\rangle\langle m|\otimes |s\rangle\langle \bm{\vec{\xi}}_{sm} | ,  \label{eq:matG}
\end{align}
and
\begin{align}
	\vec{\rho} =& \bm{\mathcal{S}}_\textsc{swap}\,\bm{\mathcal{X}}\, \bm{\mathcal{G}}^{\dashv} \,\text{vec}(\widetilde{\mathcal{P}}) \nonumber \\
		=& \bm{\mathcal{S}}_\textsc{swap}\,\bm{\mathcal{X}}\, \bm{\mathcal{G}}^{\dashv} \, \left(\mathcal{F}\otimes \mathbb{I}_{s_{\sf max}}\right) \cdot \text{vec}(\mathcal{P}), \label{eq:vec_rho_2}	
\end{align}
where $\bm{\mathcal{G}}^{\dashv}$ is the Moore-Penrose pseudoinverse matrix~\cite{Penrose1955} of~$\bm{\mathcal{G}}$, $\mathbb{I}_{s_{\sf max}}$ is a $s_{\sf max}\times s_{\sf max}$ identity matrix, and $\text{vec}(\mathcal{P})$ is the vectorization of matrix $\mathcal{P}$ (see Appendix~\ref{app:vec}). We have used Eq.~(\ref{eq:vecprod}) in order to write ${\text{vec}(\mathcal{P}\mathcal{F}) = (\mathcal{F}\otimes\mathbb{I})\,\text{vec}(\mathcal{P})}$, being~$\mathcal{F}$ a symmetric matrix. We have resorted to vectorized versions of some matrices as these allow one to write efficient numerical codes. Matrix pseudoinverse has been used instead of the usual matrix inverse because~$\bm{\mathcal{G}}$ contains~$Ds_{\sf max}$ rows and~$D^2$ columns and, consequently, may be not square. Equation~(\ref{eq:vec_rho_2}), in summary, relates the components of the reconstructed density matrix---stored in vector $\vec{\rho}$---with the experimental measurements~($\mathcal{P}$) and the measurement settings~($\bm{\mathcal{G}}$) in an explicit way. The density operator $ \rho$ is obtained by just rearranging the elements of $\vec{\rho}$.

\subsection{Stability of the inversion \label{sec:CN}}

\begin{figure}[!t]
\includegraphics[width=0.99\columnwidth]{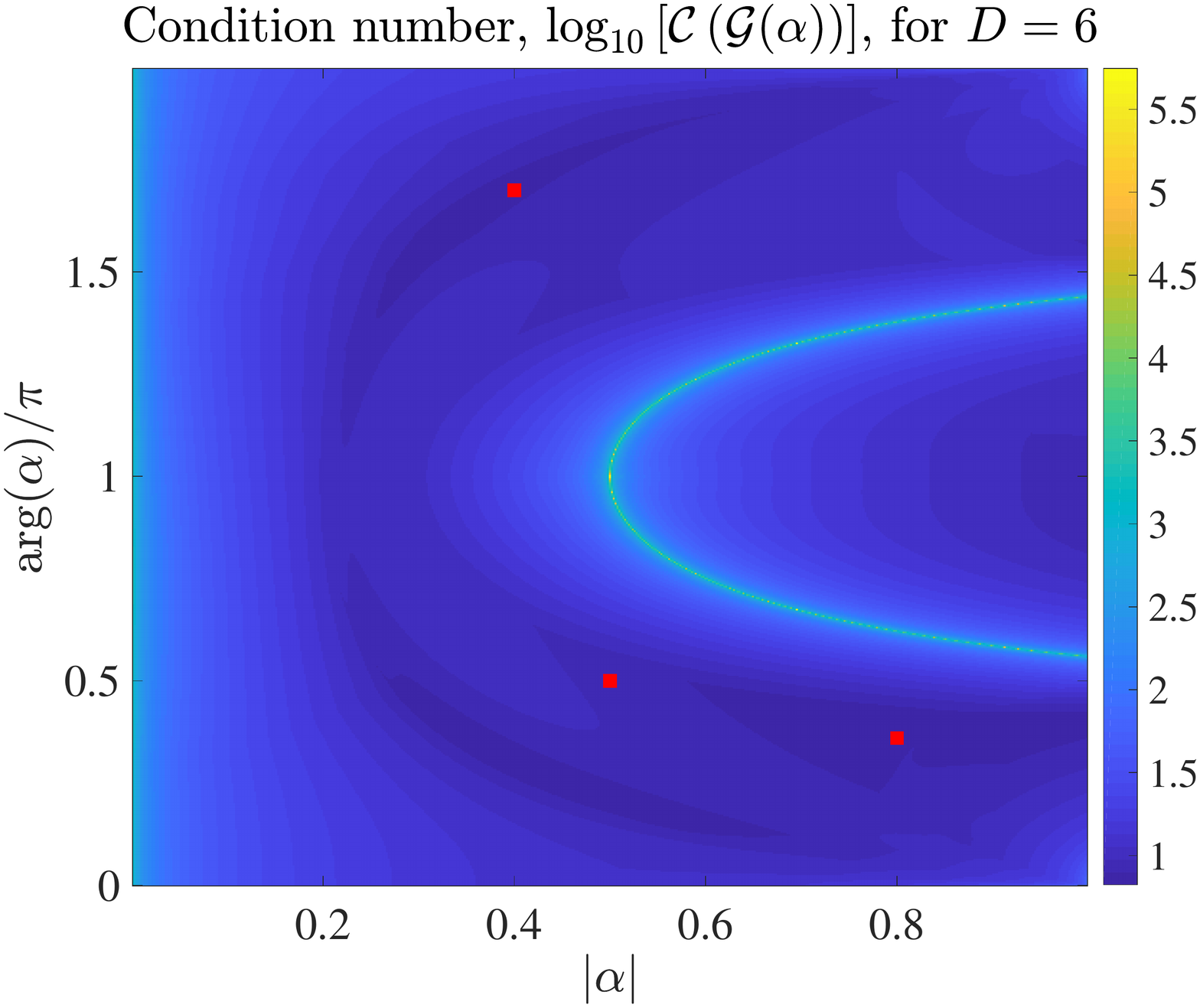}\\
\includegraphics[width=0.99\columnwidth]{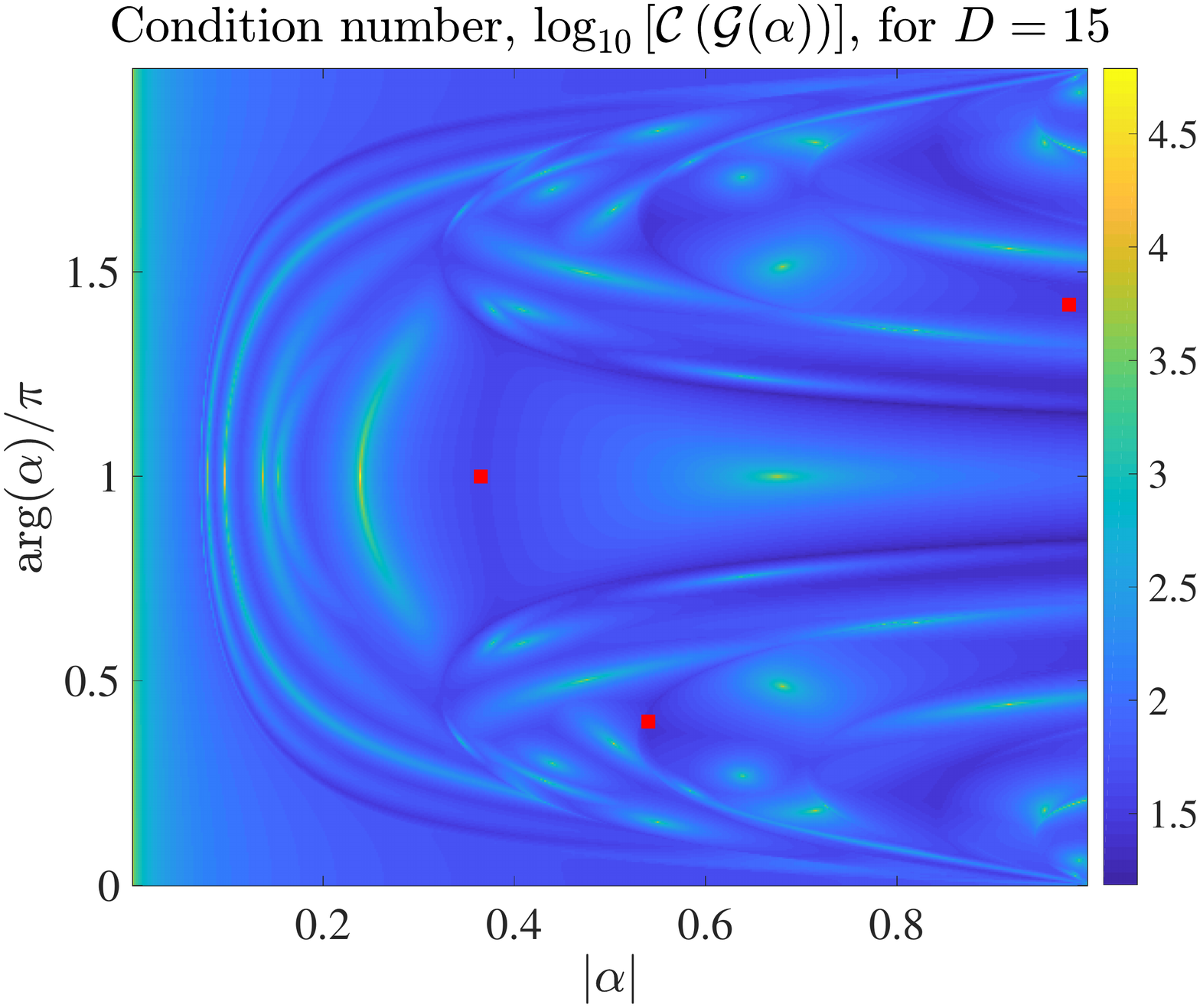}
\caption{Condition number $\mathcal{C} \left( \bm{\mathcal{G}}(\alpha) \right)$ as function of $\alpha$ given that fiducial state $|\alpha_0\rangle$ is given by Eqs.\thinspace(\ref{eq:eqidA}) and~(\ref{eq:eqidL}) for dimensions 6 and 15. Red squares indicate the values of $\alpha$ that were used in the experiment reported in this article. Since $\mathcal{C} \left( \bm{\mathcal{G}}(\alpha) \right)$ might adopt very different values, these graphs were presented in logarithmic scale.  \label{fig:CNum}}
\end{figure}

The stability of the inversion under variations of the experimentally obtainable probabilities can be studied by inspecting Eq.\thinspace(\ref{eq:vec_rho_2}). The problem is either well or ill-conditioned depending on the condition number $\mathcal{C}$ of the matrix involved in the inversion. This, in turn, depends on the singular values of such matrix~\cite{Cheney2008}. As matrices $\bm{\mathcal{S}}_\textsc{swap}$, $\bm{\mathcal{X}}$, and $\mathcal{F}\otimes\mathbb{I}_{s_{\sf max}}$ are all unitary, they do not modify singular values and, hence, matrix $\bm{\mathcal{G}}$ suffices to analyze the robustness of the tomographic procedure under experimental noise. Indeed,
\begin{align}
	\mathcal{C} \left( \bm{\mathcal{S}}_\textsc{swap}\,\bm{\mathcal{X}}\, \bm{\mathcal{G}}^{\dashv} \, \left(\mathcal{F}\otimes \mathbb{I}_{\sf s_{max}}\right) \right) = \mathcal{C} \left( \bm{\mathcal{G}} \right) = \frac{\sigma_{\max}\left( \bm{\mathcal{G}} \right)}{\sigma_{\min}\left( \bm{\mathcal{G}} \right)},\label{eq:C(G)}
\end{align}
where $\sigma_{\max}(\bm{\mathcal{G}})$ and $\sigma_{\min}(\bm{\mathcal{G}})$ stand for the maximal and minimal singular values of $\bm{\mathcal{G}}$, respectively. It can be concluded from Eqs.\thinspace(\ref{eq:Xisk}), (\ref{eq:matG}), and~(\ref{eq:C(G)}) that a study of $\mathcal{C} \left( \bm{\mathcal{G}} \right)$ as function of $|\alpha_0\rangle$ allows one to predict whether a given fiducial state will be a good choice for quantum tomography. A small condition number indicates that the fiducial state is a good candidate for building the tomographic method.

As $\mathcal{C} \left( \bm{\mathcal{G}} \right)$ depends on $D$ complex parameters, its optimization over the Hilbert space does not seem to be computationally easy. For sake of simplicity, we will resort to the notation used in Ref.~\cite{Paiva-Sanchez2010} in order to analyze~$\mathcal{C} \left( \bm{\mathcal{G}} \right)$ in terms of a single complex parameter $\alpha$. Thus, the fiducial state will be given by
\begin{align}
	|\alpha_0(\alpha)\rangle =& \sum_{k=0}^{D-1}\sqrt{\frac{\lambda_k(\alpha)}{D}}|k\rangle,\label{eq:eqidA}
\end{align}
where
\begin{align}	
	\lambda_k(\alpha) =&	1 - |\alpha|\dfrac{\sin\left( \frac{k\pi + (D-1)\arg(\alpha) }{D} \right)}{\sin\left( \frac{k\pi-\arg(\alpha) }{D} \right)}. \label{eq:eqidL}
\end{align}

Figure~\ref{fig:CNum} shows the decimal logarithm of~$\mathcal{C} \left( \bm{\mathcal{G}}\right)$ as a function of the absolute value and phase of~$\alpha$ for dimensions 6 and 15. As it can be observed, $\mathcal{C} \left( \bm{\mathcal{G}}(\alpha) \right)$ can adopt values ranging from~$\sim 10^1$ to~$\sim 10^5$, which demonstrates the necessity of a careful choice of $\alpha$ before performing the experiment. In each of the panels, three red squares highlight the values of~$\alpha$ that were used for our experiment. These values, also displayed in Table~\ref{tab:alphas}, were chosen from regions at the figures exhibiting small condition numbers. We have dealt with the problem of numerical stability by choosing fiducial states such that $\mathcal{C} \left( \bm{\mathcal{G}} \right) $ adopts small values. Instead of resorting to a given parametrization, we could have generated a large set of random fiducial states and compute the value of $\mathcal{C} \left( \bm{\mathcal{G}} \right)$ for each one. If done so, condition numbers even lower than the ones used here could be obtained. Nonetheless, using the former procedure we were able to ensure the states on a neighborhood with small condition numbers to have a more robust reconstruction in the case of having noise due to experimental imperfections.

\begin{table}[!t]
\begin{ruledtabular}
\begin{tabular}{cccc}
$D$ & $\alpha_1$ & $\alpha_2$ & $\alpha_3$ \\ \hline
$6$ & $0.4\,e^{1.7\pi i}$ & $0.8\,e^{0.36\pi i}$ & $0.5\, e^{0.5\pi i}$ \\
{} 	& $(7.796)$	& $(6.848)$ & $(8.946)$ \\
$15$ & $0.365\,e^{\pi i}$ & $0.54\,e^{0.4\pi i}$ & $0.98\,e^{1.42\pi i}$ \\
{} 	& $(40.94)$	& $(27.32)$	& $(33.15)$
\end{tabular}
\end{ruledtabular}	
\caption{\label{tab:alphas} Values of $\alpha$ chosen for experimental purposes. Numbers in parentheses below each $\alpha$ indicate the condition number, which is extracted from data of Figure~\ref{fig:CNum}.}
\end{table}

\section{Experiment \label{sec:setup}}

\begin{figure}[!h]
\includegraphics[width=\columnwidth]{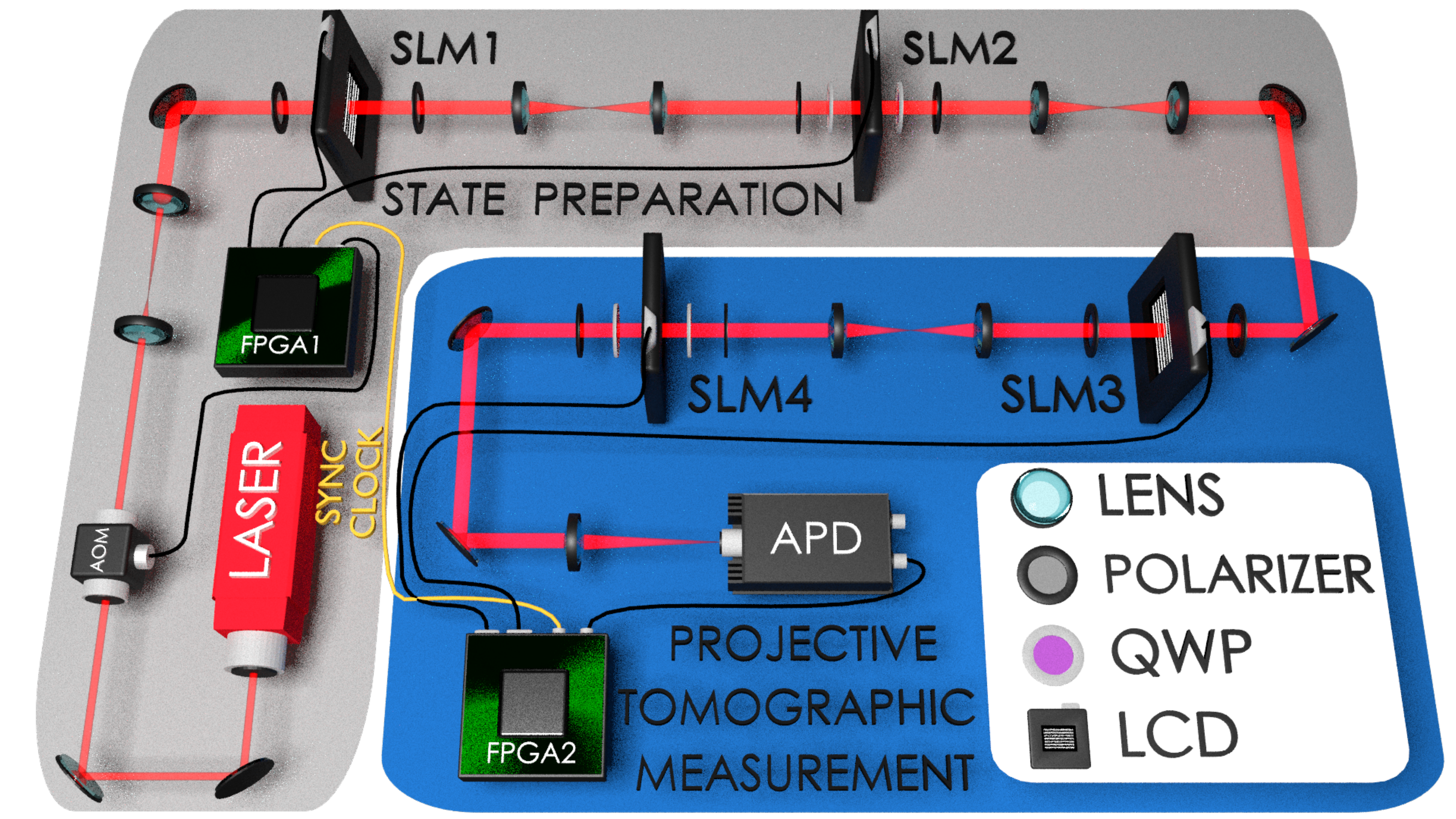}
\caption{Experimental setup. Weak coherent pulses are generated using a 690 nm continuous-wave laser, an AOM, and calibrated attenuators (not shown in figure for sake of clarity). In the SP stage, $d$ slits in SLM1 and SLM2 are used to prepare the state $|\Psi^d\ra$. In the PM stage a similar set of $d$ slits on SLM3 and SLM4, combined with a pointlike APD, implements the projection of the state $|\Psi^d\ra$ onto one of the states required in the quantum tomography protocol considered. In this way we are able to estimate $|\la\alpha_{sj}|\Psi^d\ra|^2$. \label{fig:setup}}
\end{figure}

Our setup is depicted in Fig. \ref{fig:setup}. It consists of two main blocks: the \textit{state preparation} (SP) and \textit{projective tomographic measurement} (PM) stages. In SP, weak coherent states are produced resorting to a 690 nm continuous-wave single-mode laser heavily attenuated with calibrated optical filters (not shown in Fig \ref{fig:setup} for sake of simplicity) and modulated with an Acousto-Optic Modulator (AOM) at a repetition rate of 30Hz. The mean photon number per pulse is set to $\mu=0.9$. In this case, this source works as an approximation to a nondeterministic single-photon source, since pulses with a single photon account for 62.3\% of the generated non-null pulses \cite{Gisin2002}. Contributions of multiphoton events to the recorded statistics is strongly suppressed by using a detection window much smaller than the optical pulse duration. Last, extra polarizing cubes with an overall extinction ratio greater than $10^{-7}$ are used to ensure a high quality of horizontal polarization of the transmitted photons. In this way, we are able to attain a high purity degree for the high-dimensional states generated with the spatial light modulators \cite{Torres-Ruiz2010d}.

SLMs are a central part of our setup. Each pixel of a SLM is part of a twisted nematic liquid crystal display (LCD), whose birefringence can be controlled by means of standard video signals emitted by a field programable gate array (FPGA) electronics. By properly controlling the polarization of the photon before and after crossing the LCD, we can set the SLM to work modulating only the amplitude of the light (SLM1 and SLM3) or a as a phase only modulator (SLM2 and SLM4)~\cite{Moreno2003a}. Arrays of $D$ slits are displayed on SLM1, each having a width of $96~\mu m$ and  transmittance coefficients $t_\ell$. The centers of contiguous slits are separated by $192~\mu m$. An imaging system projects the image of SLM1 on SLM2, where phases $\phi_\ell$ are added to each slit. Thus, the state of the single photons transmitted by these SLMs is $|\Psi\rangle \propto \sum^{D-1}_{\ell=0}\sqrt{t_\ell}e^{i\phi_\ell}|\ell\rangle$, and it represents a $D$-dimensional quantum system that is encoded into the linear transverse momentum of the photons \cite{Neves2005,Lima2009f,Lima2011}. $|\ell\rangle$ denotes the state of the photon transmitted by the $\ell$th-slit of the SLMs.

To test our new tomographic method we considered 3 different type of states for dimension $D=6$ and $D=15$. The reason for choosing such dimensions are: (i) to illustrate the relevance of our method while considering even and odd dimensions, (ii) the tomographic method based on mutually unbiased bases can not be used in these dimensions, and (iii) dimension $D=15$ corresponds, up to date, to the highest dimension that we have implemented a quantum state reconstruction procedure. To be more specific, the prepared states were

\begin{subequations}
\begin{align}
	\left|\Psi_1^{6}\right\rangle ~=~ & \frac{1}{\sqrt{6}} \sum_{j=0}^{5}|j\rangle, \label{eq:psi1_6}\\
	\left|\Psi_2^{6}\right\rangle ~=~ & |0\rangle,\label{eq:psi2_6}  \\
	\left|\Psi_3^{6}\right\rangle ~=~ & \frac{1}{\sqrt{6}}\Big[ \Big(|0\rangle + |2\rangle \Big) + e^{-{i\pi/4}}|4\rangle \nonumber\\ & \hspace{0.60cm}+ e^{-{i\pi/8}}\Big(|1\rangle+|3\rangle+|5\rangle\Big) \Big],\label{eq:psi3_6}
\end{align}
\end{subequations}
for dimension 6, and
\begin{subequations}
\begin{align}	
	\left|\Psi_1^{15}\right\rangle ~=~ & \frac{1}{\sqrt{15}}\sum_{j=0}^{14}|j\rangle,\label{eq:psi1_15} \\
	\left|\Psi_2^{15}\right\rangle ~=~ & |7\rangle, \label{eq:psi2_15}\\
	\left|\Psi_3^{15}\right\rangle ~=~ & \frac{1}{\sqrt{15}}\Big[ \Big(|0\rangle + |5\rangle + |8\rangle + |14\rangle \Big) \nonumber\\
		& \hspace{0.75cm} + e^{-{i\pi/10}}\Big(|1\rangle + |3\rangle + |9\rangle + |12\rangle\Big) \nonumber\\
		& \hspace{0.75cm} + e^{-{i\pi/9}}|2\rangle + e^{-{i\pi/8}}\Big(|6\rangle + |11\rangle\Big) \nonumber\\
		& \hspace{0.75cm} + e^{-{i\pi/7}}\Big(|4\rangle+|10\rangle\Big) \nonumber\\
		& \hspace{0.75cm} + e^{-{i\pi/6}}\Big(|7\rangle+|13\rangle\Big)\Big]. \label{eq:psi3_15}
\end{align}
\end{subequations}
for dimension 15.
	
The Projective Tomographic Measurement stage contains two new SLMs: SLM3 and SLM4, used for post-selecting the state to be detected. For this purpose, a new set of transmittance coefficients $\tau_\ell$ and phases $\zeta_\ell$ are used on SLM3 and SLM4, respectively. Finally, detection is performed at the center of the focal plane of a lens located after SLM4 using an avalanche single-photon detector (APD) with a $10~\mu m$-wide pinhole placed in front of it. The probability of detecting a single photon is, thus, proportional to $|\langle\Theta|\Psi\rangle|^2$~\cite{Lima2011,Etcheverry2013,Taguchi2008}, where $|\Theta\rangle\propto \sum_{\ell} \sqrt{\tau_\ell} e^{-i\zeta_\ell}|\ell\rangle$. In our case, $|\Theta\rangle$ represents each of the states $|\alpha_{sj}\rangle$ on which the measurements are performed. That is, for projecting at each $|\alpha_{sj}\rangle$, we considered different values of $\tau_\ell$ and $\zeta_\ell$. In order to show the possibility of using different fiducial states, we used $|\alpha_0(\alpha_k)\rangle$ [see Eq.~(\ref{eq:eqidA})] as the fiducial state for reconstructing state $|\Psi_k^D\rangle$, where the values of $\alpha_k$ are the ones shown in Table~\ref{tab:alphas}.

Each projective measurement related to our tomographic method was repeated 10 times, which allowed us to obtain its associated mean value of detection counts. We denote $n_{sj,r}$ as the counts obtained from measuring $\Pi_{sj}$ in the $r$th round of measurements. Average numbers of counts $\bar{n}_{sj}$ can then be obtained by
\begin{align}
	\bar{n}_{sj} = \frac{1}{10}\sum_{r=1}^{10} n_{sj,r}.
\end{align}
We considered a total of $10000\times D$ experimental runs for each round of measurements. Once all the detection counts were recorded and the average probabilities were computed, we proceeded to the post-measurement processing of the data. Error margins for density matrices and its corresponding figures of merit were determined through 10000 Monte Carlo simulations for each reconstructed state. Simulated counts numbers $n_{sj}^{(\mu)}$ are obtained by adding Poisson noise to the originally averaged data, where $\mu$ denotes the number of the Monte Carlo trial and ranges from 1 to 10000. Only in the first case there is no noise considered, i.e.,
\begin{align}
	n_{sj}^{(\mu)} =&
	\begin{cases}
		\bar{n}_{sj},	& \text{ for } \mu=1, \\
		\text{Poisson}\left( \bar{n}_{sj} \right), & \text{ otherwise.}
	\end{cases}
\end{align}

\begin{figure}[!t]
\includegraphics[width=\columnwidth]{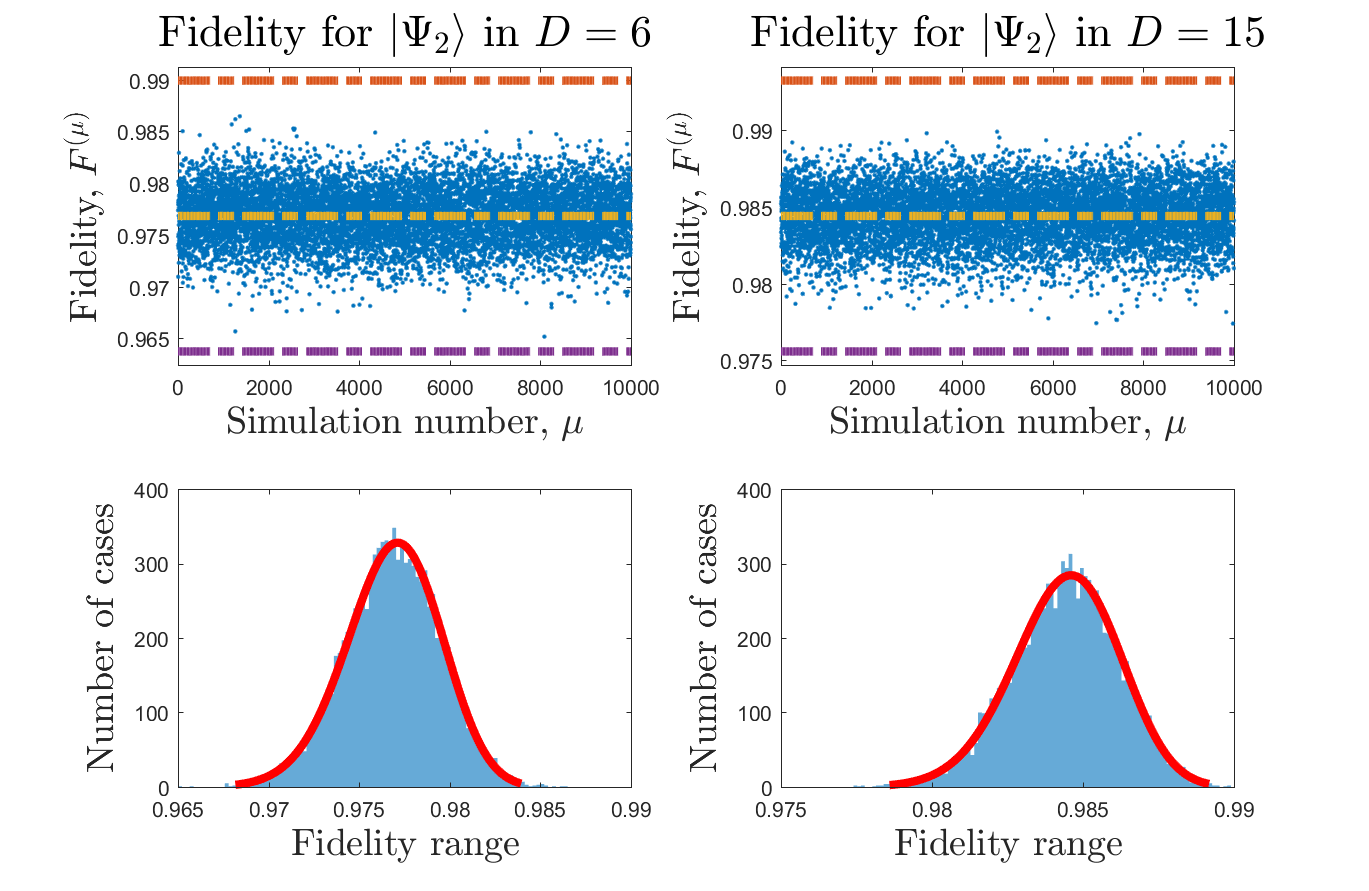}
\caption{Examples of the Monte Carlo simulations performed (uppper panels) and their respective histograms (lower panels). Three horizontal dot-dashed lines in the upper graphs represent the mean value ($\langle F\rangle$) of the simulations and the $\langle F\rangle\pm 5\sigma$ interval. The continuous line in each histogram represents a fitted beta distribution. For each fitted function, the probability of having a value outside the~$\pm 5\sigma$ interval is~$\sim 10^{-6}$.   \label{fig:hist}}
\end{figure}

\begin{figure*}
\includegraphics[width=\textwidth]{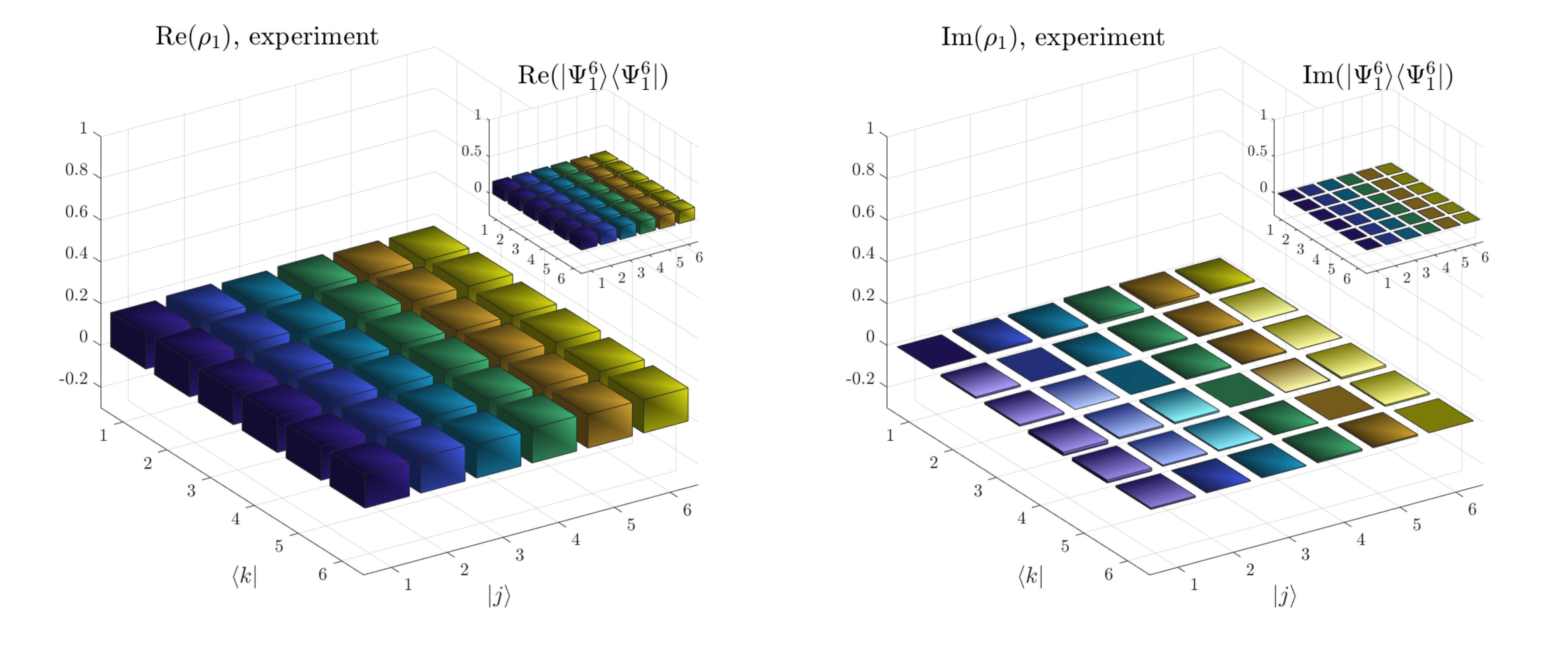}
\includegraphics[width=\textwidth]{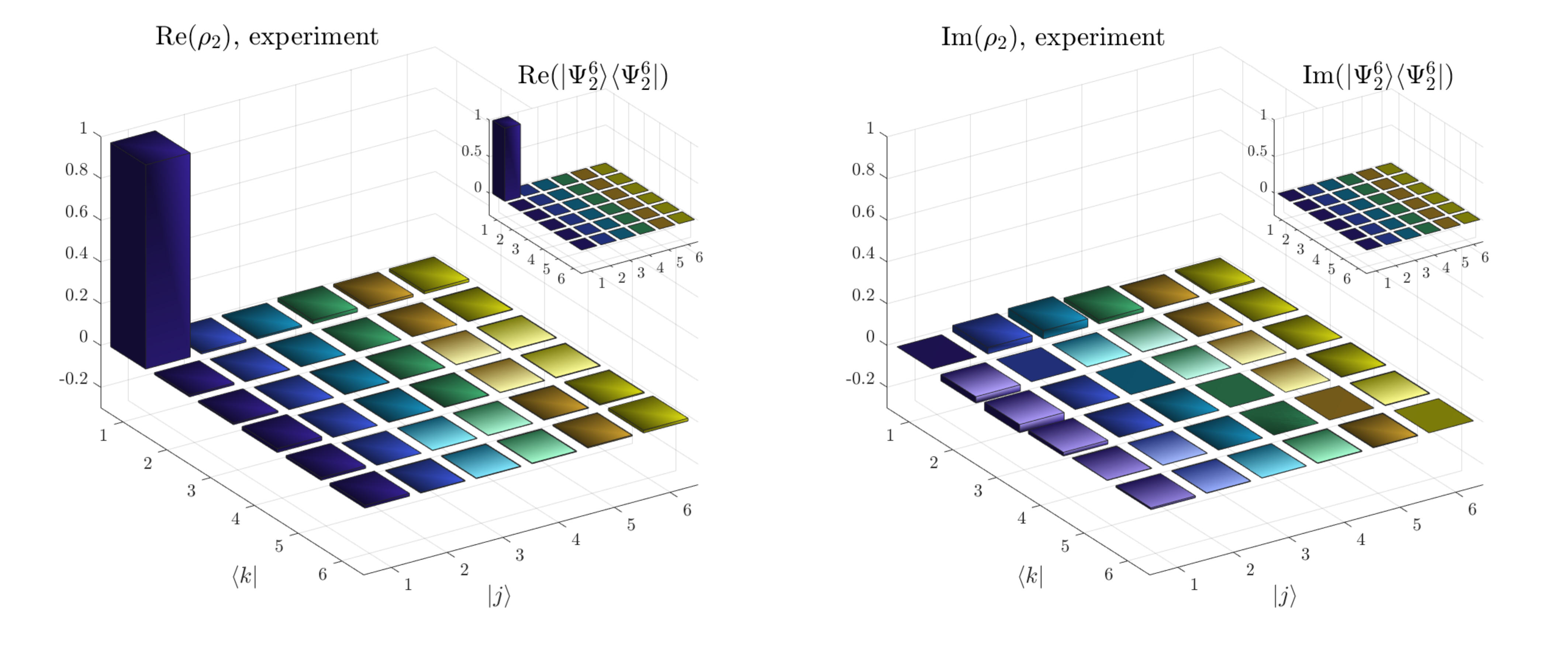}
\includegraphics[width=\textwidth]{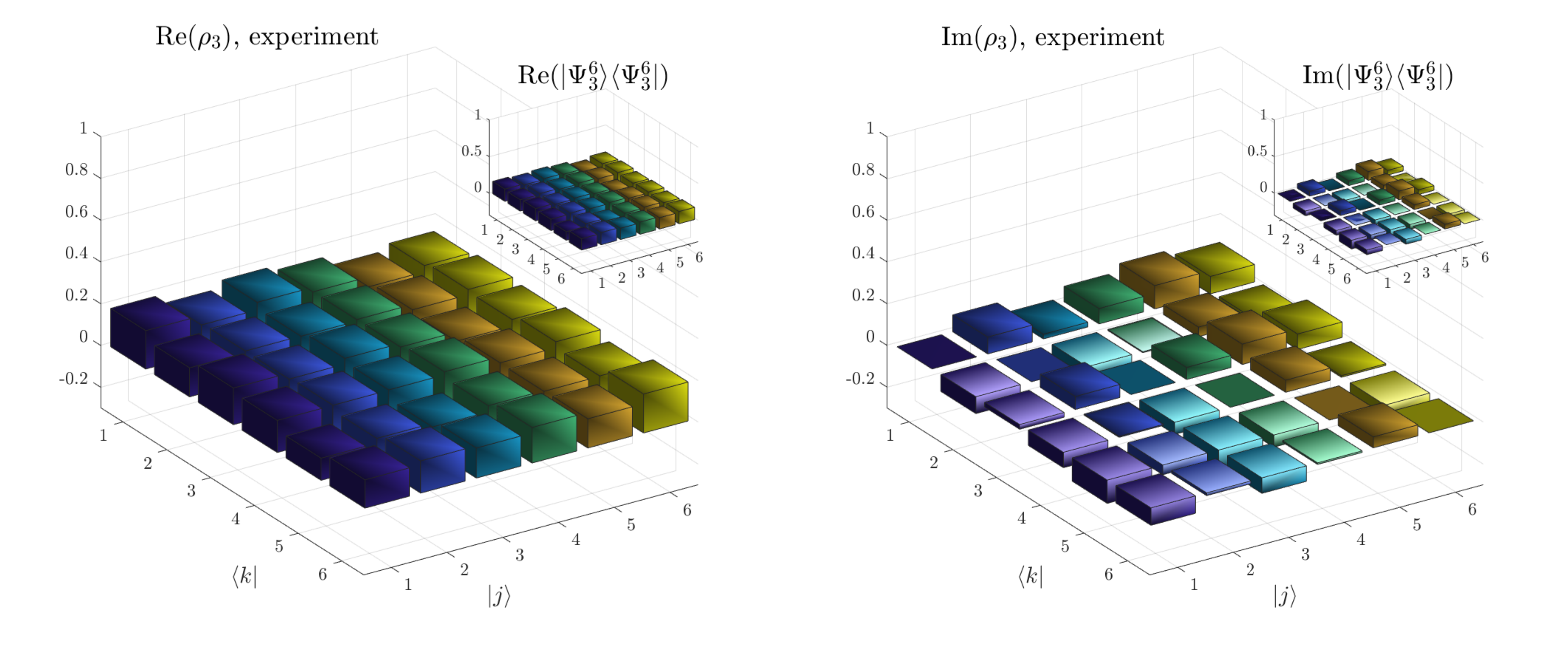}
\caption{Reconstructed quantum states for~$D=6$. The insets show the theoretically expected results. \label{fig:D06}}
\end{figure*}

\begin{figure*}
\includegraphics[width=\textwidth]{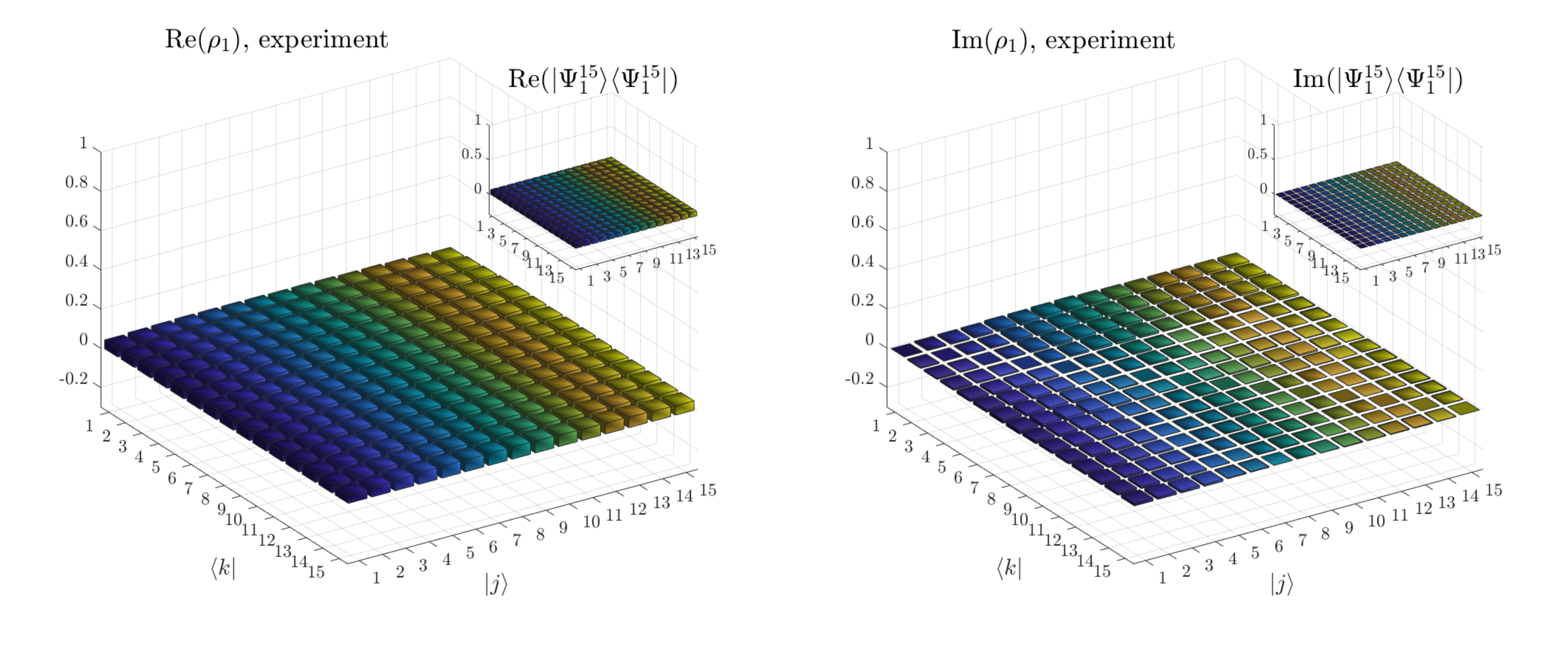}
\includegraphics[width=\textwidth]{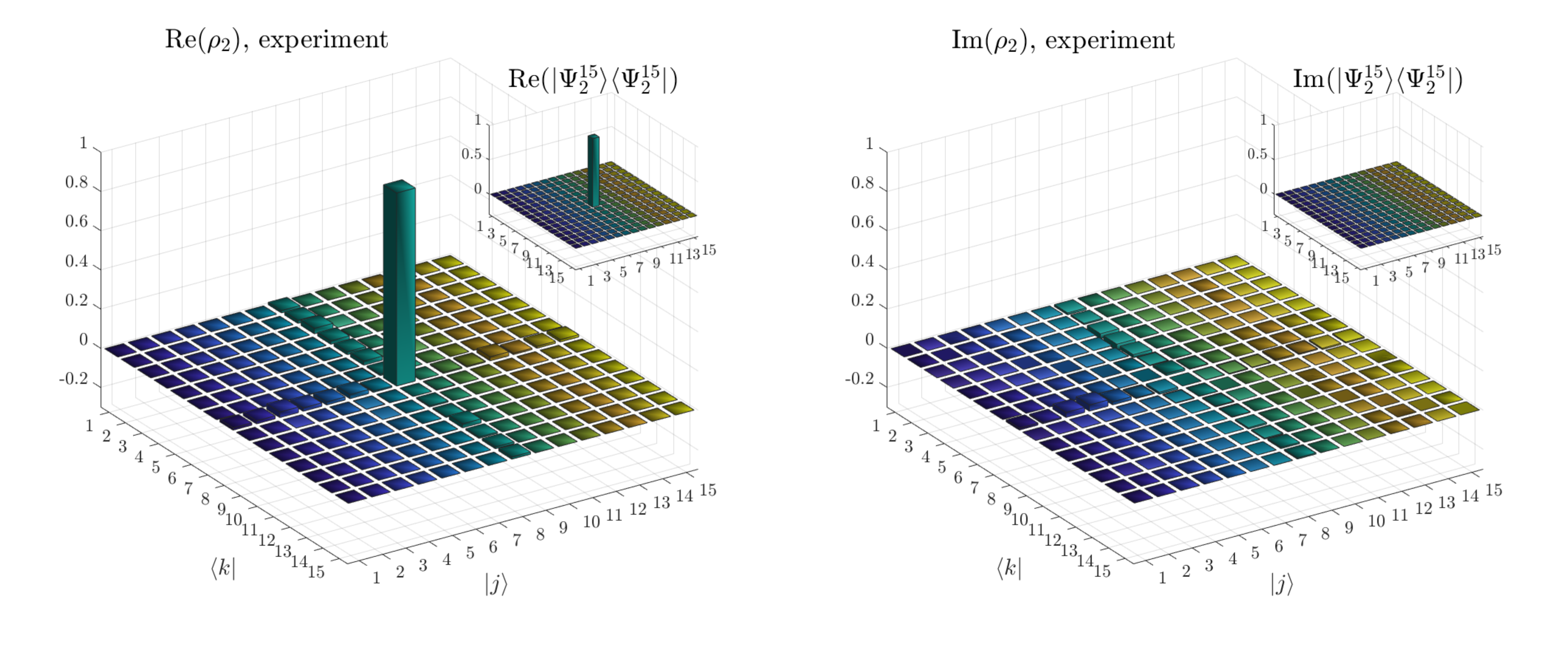}
\includegraphics[width=\textwidth]{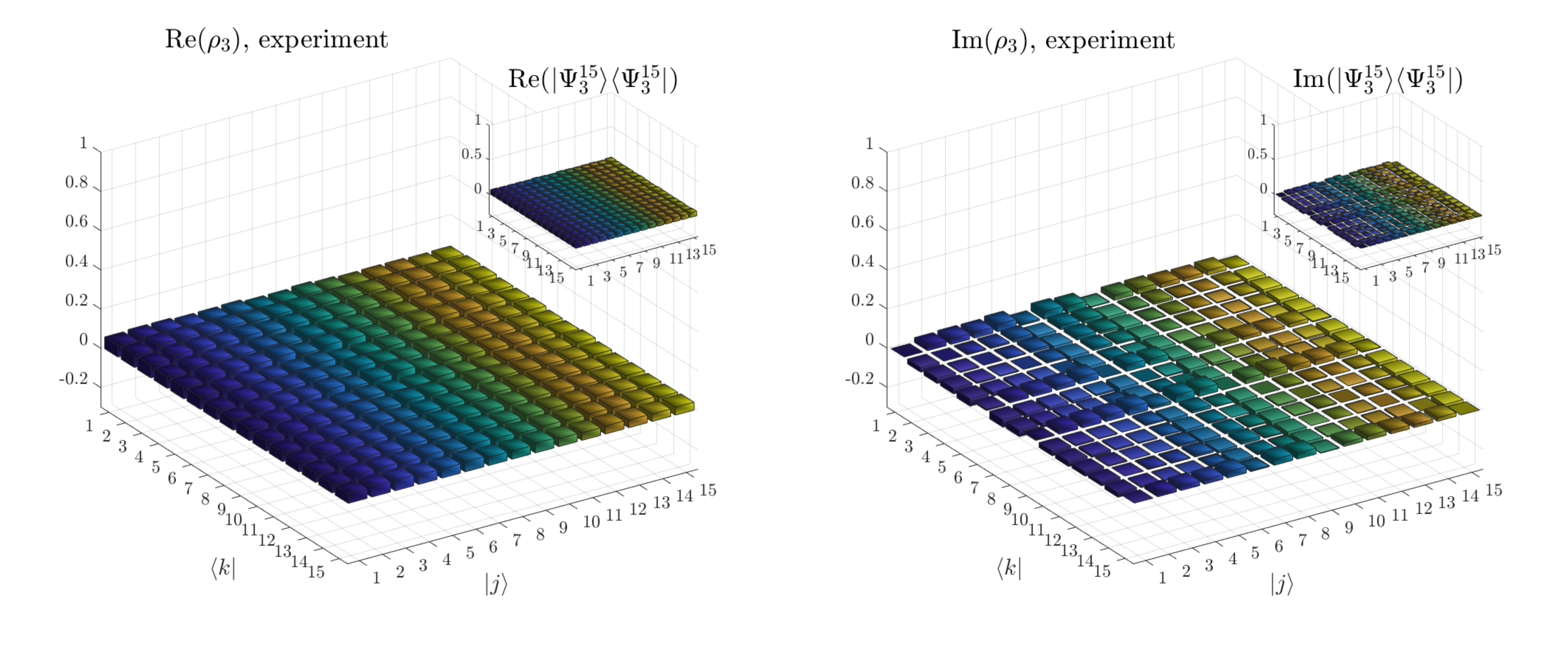}
\caption{Reconstructed quantum states for~$D=15$. The insets show the theoretically expected results. \label{fig:D15}}
\end{figure*}

Afterwards, simulated probability matrices $\mathcal{P}^{(\mu)}$ are computed for each Monte Carlo trial according to
\begin{align}
	\mathcal{P}^{(\mu)} =& \dfrac{\displaystyle\sum_{s=0}^{s_{\sf max}-1}\sum_{j=0}^{D-1} n_{sj}^{(\mu)} |s\rangle\langle j|}{\displaystyle\sum_{t=0}^{s_{\sf max}-1}\sum_{k=0}^{D-1} n_{tk}^{(\mu)}},
\end{align}
where Eq.\thinspace(\ref{eq:POVM}) was taken into account. Then, matrix $\mathcal{P}^{(\mu)}$ is used in Eq.\thinspace(\ref{eq:vec_rho_2}) in order to obtain a reconstructed density matrix $\rho^{(\mu)}$. As Eq.\thinspace(\ref{eq:vec_rho_2}) cannot ensure its positiveness, maximum likelihood estimation (MLE) was subsequently employed (see Appendix~\ref{app:MLE} for details) in order to ensure matrix positiveness~\cite{Hradil1997,James2001,Thew2002}. The fidelity $F^{(\mu)}$ between $\rho^{(\mu)}$ and the state $|\Psi\rangle$ we intended to prepare is computed as figure of merit for each state resulting from MLE, where
\begin{align}
	F^{(\mu)} =& \langle\Psi| \rho^{(\mu)} |\Psi\rangle.
\end{align}
The final result for the fidelity is expressed in terms of the mean and standard deviation of the simulated results, that is,
\begin{align}
	F =& \left\langle \left\{ F^{(\mu)} \right\}\right\rangle \pm 5\sigma\left( \left\{ F^{(\mu)} \right\}\right).
\end{align}
Two examples of the Monte Carlo simulations are shown in Fig.\thinspace\ref{fig:hist}. We have chosen $\pm 5\sigma$ as error margins since the probability of obtaining a value outside it in a new round of experiments is less than $10^{-6}$ in the case the values of $F^{(\mu)}$ distribute around their mean value following a normal distribution. In the worst-case scenario, such probability is less than $4\%$, according to the Bienaym\'e-Chebyshev inequality. The reconstructed density operators in dimension 6 are depicted in Fig.\thinspace\ref{fig:D06}, whereas Fig.\thinspace\ref{fig:D15} illustrate the results for dimension 15.  A summary of the results is shown in Table~\ref{tab:fids}. As it can be seen, higher values of fidelities were obtained. More specifically, for dimension 6 (15) an overall fidelity of 0.977 (0.957) has been recorded, while considering an ensemble of only $6\times10^4$ ($1.5\times10^5$) events of photo-detection for the state reconstruction procedure. Other similar optical setups validates the good performance of the method presented here: Ref.~\cite{Lima2011} reported the experimental realization of tomography using mutually unbiased bases and they obtained fidelities of $0.96\pm0.03$ and  $0.93\pm0.03$ for dimensions $7$ and $8$, respectively. Ref.~\cite{Goyeneche2015} reported $0.985\pm0.015$ for $D=8$ using a method designed for reconstructing pure states. The experiment of Ref.~\cite{Bent2015} using SIC-POVM obtained fidelities of $0.960\pm0.003$ and $0.887\pm0.003$ for dimensions 6 and 10, respectively.

\begin{table}[!t]
\begin{ruledtabular}
\begin{tabular}{cccc}
$D$ & $\left|\Psi_1^{D}\right\rangle$ & $\left|\Psi_2^{D}\right\rangle$ & $\left|\Psi_3^{D}\right\rangle$ \\ \hline
	 $6$ & $0.998\pm0.001$ & $0.977\pm0.013$ & $0.956\pm0.010$\\
	$15$ & $0.965\pm0.006$ & $0.984\pm0.009$ & $0.922\pm0.007$
\end{tabular}
\end{ruledtabular}	
\caption{\label{tab:fids} Fidelities obtained for each of the reconstructed states, with their respective $5\sigma$ uncertainty extracted from 10000 Monte Carlo trials. MLE was used in each trial.
}
\end{table}

\section{Concluding remarks \label{sec:conclusion}}

In summary, we have reported the experimental realization of quantum state tomography assisted by multiply symmetric states for dimensions $D=6$ and $D=15$. Unlike MUB and SIC-POVM tomographic methods, this method is guaranteed to exist in any dimension and provides a significant reduction in the number of measurement outcomes when compared to standard quantum tomography. Furthermore, in the case of odd dimensions the method requires the least possible number of measurement outcomes. The tomographic method is different from an arbitrary linear inversion in the sense that multiply symmetric states constitute an informationally complete set of measurements. As Eq.~(\ref{eq:multisym2}) shows, these states depend on a given fiducial state $|\alpha_0\rangle$ that can be freely chosen as we experimentally demonstrated. Nevertheless, this fiducial state is chosen in such a way the inversion algorithm remains stable. The stability can be analyzed in terms of the condition number of matrix $\bm{\mathcal{G}}$ of Eq.\thinspace(\ref{eq:matG}).

Further improvements can be obtained by studying the condition number. We have reduced the complexity of this problem by studying fiducial states defined by two parameters, which led to condition numbers of the order of 10. However, Monte Carlo simulations with randomly generated fiducial states have shown that smaller condition numbers are possible. Recently, it has been demonstrated that two-stage quantum tomography leads to a quadratic improvement in the accuracy of the estimation of pure states of high-dimensional quantum systems~\cite{Pereira2018}. In the first stage of this adaptive tomographic method a small ensemble is employed to obtain a first estimate via standard quantum tomography. This estimate's eigenstates are employed to represent the generators of SU($D$), which are subsequently measured in a second stage of standard quantum tomography. Analogously, we can consider an adaptive version of quantum tomography assisted by multiply symmetric states. Since this method requires less measurements than standard quantum tomography, it seems that for a given ensemble size, a higher accuracy might be achieved. Other continuation of the current work concerns the case of multipartite systems. For instance, the state of a two-qudit system can be can be estimated with a minimal number of $D^4$ measurement outcomes. This can be achieved for $D$ odd by conditional local estimations employing quantum tomography assisted by multiply symmetric states.

\begin{acknowledgments}
This work was supported by CONICYT FONDECYT 1160400, 3170400, 11150324, 11121318,  1180558, and Millennium Institute for Research in Optics (MIRO). D.\thinspace M. acknowledges financial support from CONICYT Doctorado Nacional~2116050.
\end{acknowledgments}

\appendix

\section{Matrix vectorization~\label{app:vec}}

Let us consider a general $M\times N$ matrix $A=\sum_{kl}A_{kl}|k\rangle\langle l|$ written in the computational basis. Its vectorization $\text{vec}(A)$ is obtained from a linear operation such that
\begin{align}
	\textrm{vec}:\mathbb{C}^{M\times N} &\rightarrow \mathbb{C}^{N}\otimes\mathbb{C}^{M}\nonumber \\
									A &\mapsto \textrm{vec}(A) = \sum_{kl}A_{kl}|l\rangle\otimes|k\rangle,
\end{align}
where vectors $\{|j\rangle\}_{j=1}^{n}$ correspond to the computational basis. Vector $\textrm{vec}(A)$ contains the columns of $A$ stacked one on top of another. Hence, we can also write a matrix vectorization as
\begin{align}
\text{vec}(A) = \sum_{j=1}^{N} |j\rangle\otimes A|j\rangle. \label{eq:vectKron}
\end{align}
Additional properties of vectorization are
\begin{align}
	\text{vec}(AB)  =& (\mathbb{I} \otimes A)\text{vec}(B) = (B^\intercal \otimes\mathbb{I} )\text{vec}(A), \label{eq:vecprod}\\
	{\rm tr}(A^\dagger B) = & \langle \text{vec}(A) | \text{vec}(B) \rangle, \label{eq:vecbraket}
\end{align}
	Faster computations of probabilities can be performed by resorting to matrix vectorization. Indeed, if we need to compute a vector given by ${\mathbf{p} =  \sum_{\mu} {\rm tr}(\Pi_\mu \rho)|\mu \rangle}$, where $\Pi_\mu$ are hermitian operators, then
\begin{align}
	\mathbf{p} =&  \sum_{\mu} {\rm tr}(\Pi_\mu \rho)|\mu \rangle \nonumber \\
		\overset{(\ref{eq:vecbraket})}{=}&  \sum_{\mu } \langle \text{vec}(\Pi_\mu) | \text{vec}(\rho) \rangle|\mu \rangle \nonumber \\
		=&  \left(\sum_{\mu } |\mu \rangle \langle \text{vec}(\Pi_\mu) | \right)  | \text{vec}(\rho) \rangle \nonumber \\
		=& \bm{\hat{\Pi}}^{\dagger}|\text{vec}(\rho)\rangle, \label{eq:pVector}
\end{align}
where $\bm{\hat{\Pi}}$ is a matrix whose $\mu$th column is the vectorization of $\Pi_\mu$. Equation~(\ref{eq:pVector}) is very useful when probabilities must be computed a large number of times from a constant set of probability operators.

\section{Efficient computation of MLE based on Poisson distribution \label{app:MLE}}
We define $N_j^\text{th}(\varrho)=\eta_j {\rm tr}(\Pi_j \varrho) + d_j$ as the theoretically expected number of counts for the $j$th detector subject to detection efficiency $\eta_j$ and a mean number of dark counts given by $d_j$. Matrix $\varrho$ is a positive operator representing an \emph{unnormalized} density matrix whose trace serves as a mean ensemble size. Additionally, $\bm{n}$ and $\bm{N}_\text{th}$ will be vectors containing the experimental and theoretical number of counts, respectively. Vector $\bm{N}_\text{th}(\varrho)$ can be efficiently written, with aid of Equation~(\ref{eq:pVector}), as
\begin{align}
	\bm{N}_\text{th}(\varrho) 
			=& \bm{\eta}\circ \left(\widehat{\bm{\Pi}}^{\dagger} \bm{\vec{\varrho}} \right) + \bm{d}, \label{eq:nth}
\end{align}
where
\begin{align}
	&\bm{\eta} =  \sum_{j\in\mathcal{M}} \eta_j |j\rangle,  &	
	&\bm{d} =  \sum_{j\in\mathcal{M}} d_j |j\rangle, \\
	&\bm{\vec{\varrho}} = |\text{vec}(\varrho)\rangle, 	&
	&\bm{\hat{\Pi}} = \sum_{j\in\mathcal{M}} |\text{vec}(\Pi_j) \rangle \langle j|. \label{eq:varrhoPi}
\end{align}
In case of composite systems, computation of $\widehat{\bm{\Pi}}^{\dagger} \bm{\vec{\varrho}}$ can be performed very efficiently by following the methods used in Ref.~\cite{Shang2017}.

As the number of photons provided by the source cannot be ensured to be equal for each measurement, we may resort to Poisson statistics, where the number of observed events will be given by $n_j$ and the expected mean number of events is $N_j^\text{th}(\varrho)$. Thus, the joint probability of having $n_1$ counts in detector 1 \emph{and} $n_2$ counts in detector 2 \emph{and} so and so, \emph{given that} the ensemble state is $\varrho$, can be expressed as
\begin{align}
	\mathcal{L}_\text{P}(\varrho) = & \prod_{j\in\mathcal{M}} \dfrac{e^{-N_j^\text{th}(\varrho)} \left[N_j^\text{th}(\varrho)\right]^{n_j} }{n_j !}, \label{eq:LP} 
\end{align}
where $\mathcal{M}$ is the set of measurements labels. Instead of working with $\mathcal{L}_\text{P}(\varrho)$, it is highly recommended to make use of the negative log-likelihood, ${L_\text{P}(\varrho) = -\ln \mathcal{L}_\text{P}(\varrho)}$, since maximization of $\mathcal{L}_\text{P}(\varrho)$ is equivalent to minimization of~$L_\text{P}(\varrho)$~\cite{Banaszek1999}. Negative log-likelihood is, thus, given by
\begin{align}
	L_\text{P}(\varrho) = & \sum_{j\in\mathcal{M}} \left[ N_j^\text{th}(\varrho) - {n_j}\ln \left(N_j^\text{th}(\varrho)\right)  + \ln\Gamma(n_j +1)\right] \nonumber \\
	= & \bm{u}^\intercal \bm{N}_\text{th}(\varrho) - \bm{n}^\intercal \ln\Big(\bm{N}_\text{th}(\varrho)\Big)  + \bm{u}^\intercal \ln\Big(\Gamma(\bm{n} +1) \Big), \label{eq:logLP}
\end{align}
where $\bm{u}$ is a column vector whose entries are all equal to~1 and the logarithm of a vector is used as $[\ln(\bm{a})]_j = \ln([\bm{a}]_j)$. By using Equations~(\ref{eq:nth})-(\ref{eq:varrhoPi}), the negative log-likelihoods of Equation~(\ref{eq:logLP}) can be efficiently computed. It is noteworthy that $L_\text{P}(\varrho)$ is a convex function defined over the convex set of positive operators. Consequently, the tools of convex optimization can be used to find its minima, $L_\text{P}(\varrho_\text{opt})$. The estimated density operator $\rho_\text{opt}$ is obtained by normalizing $\varrho_\text{opt}$ after the optimization has finished.

\bibliographystyle{apsrev4-1}
%

\end{document}